%% file: main_draft.tex
\documentclass[11pt]{article}
\usepackage{amssymb}
\usepackage{amsmath}
\usepackage{natbib}
\usepackage{algorithm}
\usepackage{algorithmic}
\usepackage{adjustbox}
\usepackage{multirow}
\usepackage{booktabs}
\usepackage{subcaption}
\usepackage{authblk}

\begin{document}

\title{Adaptive Gate-Aware Mamba Networks for Magnetic Resonance Fingerprinting}

\author[1]{Tianyi Ding}
\author[1]{Hongli Chen}
\author[2,1]{Yang Gao}
\author[3]{Zhuang Xiong}
\author[1]{Feng Liu}
\author[4,5]{Martijn A. Cloos}
\author[6,1]{Hongfu Sun\thanks{Corresponding author. Email: hongfu.sun@newcastle.edu.au}}

\date{}

\affil[1]{School of Electrical Engineering and Computer Science, University of Queensland, Brisbane, QLD 4072, Australia}
\affil[2]{School of Computer Science and Engineering, Central South University, Changsha 410083, China}
\affil[3]{Image X Institute, Sydney School of Health Sciences, Faculty of Medicine and Health, University of Sydney, Sydney, NSW 2006, Australia}
\affil[4]{Donders Institute for Brain Cognition and Behaviour, Radboud University, Nijmegen, Netherlands}
\affil[5]{Australian Institute for Bioengineering and Nanotechnology, University of Queensland, St Lucia, QLD, Australia}
\affil[6]{School of Engineering, University of Newcastle, Callaghan, NSW 2308, Australia}

\maketitle

\begin{abstract}
Magnetic Resonance Fingerprinting (MRF) enables fast quantitative imaging by matching signal evolutions to a predefined dictionary. However, conventional dictionary matching suffers from exponential growth in computational cost and memory usage as the number of parameters increases, limiting its scalability to multi-parametric mapping.
To address this, recent work has explored deep learning-based approaches as alternatives to DM. We propose \textbf{GAST-Mamba}, an end-to-end framework that combines a dual Mamba-based encoder with a Gate-Aware Spatial-Temporal (GAST) processor. Built on structured state-space models, our architecture efficiently captures long-range spatial dependencies with linear complexity.
On 5× accelerated simulated MRF data (200 frames), GAST-Mamba achieved a T1 PSNR of 33.12~dB, outperforming SCQ (31.69~dB). For T2 mapping, it reached a PSNR of 30.62~dB and SSIM of 0.9124. In vivo experiments further demonstrated improved anatomical detail and reduced artifacts. Ablation studies confirmed that each component contributes to performance, with the GAST module being particularly important under strong undersampling.
These results demonstrate the effectiveness of GAST-Mamba for accurate and robust reconstruction from highly undersampled MRF acquisitions, offering a scalable alternative to traditional DM-based methods.
\end{abstract}


\section{Introduction}
\label{sec1}
Magnetic Resonance Fingerprinting (MRF) has significantly advanced the ability to quantitatively characterize human brain tissues \citep{ma2013magnetic,bipin2019magnetic}. One of its key advantages is the capability to simultaneously generate multiple quantitative maps (such as T1 and T2) from a single, short scan, effectively addressing critical limitations of conventional quantitative MRI (qMRI) techniques \citep{gomez2020rapid,stikov2015accuracy,deoni2005high}. The original MRF framework introduced a three-step procedure to estimate tissue parameters \citep{coronado2021spatial}. First, a transient-state acquisition is performed by varying sequence parameters such as repetition time (TR), flip angle (FA), and k-space sampling, resulting in a unique temporal signal evolution for each voxel that serves as a fingerprint of the underlying tissue properties. Second, a signal dictionary is generated by simulating signal evolutions over a predefined range of parameter combinations using the Bloch equation\citep{Bloch1946equ}. Finally, quantitative parameter maps are estimated by matching each voxel’s measured signal trajectory to the closest entry in the dictionary using an inner product-based similarity metric \citep{ma2013magnetic,coronado2021spatial}. However, the size of the precomputed dictionary grows exponentially with the number of parameters, making the pattern-matching step increasingly challenging due to the trade-off between computational efficiency and reconstruction accuracy \citep{chen2022technical}.

To address these inherent bottlenecks, a variety of methods have been proposed. Broadly, existing efforts can be categorized into two main directions: reducing the complexity of the dictionary matching process, and replacing it entirely with direct signal-to-parameter mapping.

The first direction focuses on accelerating dictionary matching by reducing the dimensionality of both the signal and the dictionary. Several techniques have been developed for this purpose. For example, a singular value decomposition (SVD)-based method projects both the dictionary and voxel-wise signals into a low-dimensional temporal subspace, enabling more efficient matching \citep{mcgivney2014svd}. A related approach applies principal component analysis (PCA) to group highly correlated dictionary entries, significantly reducing the search space \citep{cauley2015fast}. Building on these ideas, randomized SVD has been employed to further reduce memory usage and computation time by directly estimating the dictionary’s low-rank structure \citep{yang2018low}. While these methods reduce computational complexity, performing dictionary matching in a reduced subspace may increase the risk of information loss during signal comparison, potentially affecting the fidelity of parameter estimation.

The second direction addresses the limitations of dictionary matching by directly learning a mapping from the temporal MRF signals to quantitative parameter maps using data-driven deep learning approaches. Deep neural networks inherently good at modeling complex nonlinear relationships \citep{zhang2023dive}, making them well suited for mapping temporal MRF signals to quantitative parameter maps \citep{McGivney2020review,poorman2020magnetic, gao2022instant}. This data-driven approach offers both significant computational speed-ups and improved reconstruction accuracy compared to conventional dictionary matching (DM) methods. 

Early deep learning-based MRF reconstruction approaches primarily employed fully connected neural networks due to their simplicity and computational efficiency. A representative method, DRONE \citep{cohen2018mr}, utilized a four-layer fully connected architecture, achieving reconstruction speeds up to thousands of times faster than traditional DM. However, these early models faced a critical limitation: by independently processing each voxel, they failed to incorporate spatial context, leading to reduced accuracy and consistency in reconstructed parameter maps. Subsequent studies \citep{golbabaee2019geometry, hamilton2021deep, martinez2024phase} aimed to enhance these models by modifying network structures, but the fundamental issue of neglecting spatial information persisted, particularly in highly undersampled acquisitions.

To address this issue, convolutional neural networks (CNNs) were introduced to leverage spatial coherence. The SCQ method \citep{fang2019deep}, which combined a fully connected layer for temporal processing and a U-Net architecture for spatial refinement, significantly improved reconstruction accuracy under accelerated acquisitions. Building upon this, the MRF-Mixer \citep{ding2025mrf} further advanced CNN-based approaches by integrating multi-branch spatial processing paths, demonstrating that single-shot undersampled data reconstructed using CNNs could match the accuracy typically achievable only by multi-shot acquisitions. Nonetheless, CNN-based methods remained constrained by their limited receptive fields, hindering their ability to effectively capture global spatial dependencies.

More recently, attention-based architectures have emerged in MRF to address the CNNs’ limitations regarding global spatial modeling. The CONV-ICA model \citep{soyak2021channel} introduced channel-wise attention mechanisms into convolutional networks to enhance feature selection and improve reconstruction performance. Additionally, Transformer-based architectures, such as the Local and Global Vision Transformer (LG-ViT) \citep{li2024deep}, further pushed the boundary by incorporating hierarchical spatial self-attention mechanisms, achieving notable improvements in modeling global spatial dependencies and consequently better reconstruction outcomes. However, despite these advances, Transformer-based architectures suffer from high computational costs and training instability, particularly when applied to high-dimensional spatiotemporal data such as MRF. This limits their scalability in practical settings.

A key challenge in existing architectures lies in capturing long-range spatial dependencies. While CNNs leverage local convolutions, their effective receptive field remains limited, making it difficult to model global spatial consistency. Transformer-based models expand the receptive field via self-attention mechanisms but do so at the cost of quadratic complexity in input size. To overcome these limitations, recent work in sequence modeling has introduced Mamba and its variants \citep{gu2023mamba,guo2025mambair,wang2024mamba,yue2024medmamba,xing2024segmamba} as efficient alternatives to Transformers. Mamba leverages structured state space models (SSMs) and a selective scanning mechanism to capture long-range dependencies with linear time complexity (\(O(N)\)). These characteristics make Mamba particularly suitable for modeling high-resolution spatiotemporal data, where both global spatial structure and temporal evolution must be efficiently captured.

In MRF, each voxel is characterized by a temporal fingerprint, and accurate reconstruction requires capturing both its temporal dynamics and spatial dependencies introduced by severe undersampling and the resulting image aliasing. CNN-based approaches often suffer from limited receptive fields, while Transformer models incur substantial computational overhead. To overcome these limitations, we propose GAST-Mamba, a lightweight framework that integrates Mamba-based state space modeling with adaptive spatial feature enhancement. Mamba enables efficient modeling of long-range temporal and spatial patterns with linear complexity, making it well-suited for highly undersampled MRF scenarios.

A central component of our design is the Gate-Aware Spatial-Temporal Adaptive (GAST) module, which improves spatial coherence by selectively combining spatial textures and temporal context through multi-scale gated processing. To the best of our knowledge, this is the first work that applies Mamba-based state space modeling to MRF reconstruction, offering a scalable and efficient solution tailored for highly undersampled quantitative imaging.

\section{Materials and Methods}
\label{sec:Materials and Methods}

\subsection{Preliminaries}
\subsubsection{Deep Learning Formulation of MRF Reconstruction}
MRF aims to estimate voxel-wise tissue parameters, such as T1 and T2, from temporal signal evolutions acquired under varying acquisition parameters \citep{ma2013magnetic}. Let $\mathbf{s}_i \in \mathbb{C}^t$ denote the fingerprint at voxel $i$, where $t$ is the number of time points. Deep learning-based methods typically learn a nonlinear mapping from the input signal to quantitative parameters using a neural network $f_\theta$:
\begin{equation}
\hat{\mathbf{p}}_i = f_\theta(\mathbf{s}_i), \quad \hat{\mathbf{p}}_i = \{T1, T2\} .
\end{equation}

The network is trained to approximate the inverse mapping from signal evolution to tissue parameters, either using simulated dictionary data or in vivo data with reference labels. Depending on the design, $f_\theta$ may process each voxel independently or incorporate spatial context via convolutional or attention-based operations.

\subsubsection{Structured State Space Models}

Structured State Space Models (SSMs) \citep{gu2020hippo,yue2024medmamba,wang2024lkm} offer a principled approach for modeling sequential data through linear time-invariant systems. A classical SSM describes the evolution of a latent state $h(t) \in \mathbb{R}^H$ under an external input $x(t) \in \mathbb{R}^L$ via the following ordinary differential equations:
\begin{equation}
h'(t) = \mathbf{A} h(t) + \mathbf{B} x(t), \quad y(t) = \mathbf{C} h(t),
\end{equation}
where $\mathbf{A} \in \mathbb{R}^{H \times H}$ models the state transition, and $\mathbf{B} \in \mathbb{R}^{H \times 1}$ and $\mathbf{C} \in \mathbb{R}^{1 \times H}$ represent input-to-state and state-to-output mappings.

To use SSMs in deep learning pipelines, the continuous system is discretized. Using the zero-order hold method with step size $\Delta$, the system matrices become:
\begin{equation}
\bar{\mathbf{A}} = \exp(\Delta \mathbf{A}), \quad 
\bar{\mathbf{B}} = \left(\Delta \mathbf{A}\right)^{-1} \left(\exp(\Delta \mathbf{A}) - \mathbf{I}\right) \Delta \mathbf{B},
\end{equation}
where $\mathbf{I}$ is the identity matrix. This yields the discrete-time system:
\begin{equation}
h(k) = \bar{\mathbf{A}} h(k-1) + \bar{\mathbf{B}} x(k), \quad y(k) = \mathbf{C} h(k).
\end{equation}

Unrolling the recurrence leads to a convolutional representation of the output:
\begin{equation}
y = x * \bar{\mathbf{K}}, \quad \bar{\mathbf{K}} = \left( \mathbf{C} \bar{\mathbf{B}},\ \mathbf{C} \bar{\mathbf{A}} \bar{\mathbf{B}},\ \dots,\ \mathbf{C} \bar{\mathbf{A}}^{L-1} \bar{\mathbf{B}} \right),
\label{eq:mamba-state}
\end{equation}
where $*$ denotes 1D convolution over the input sequence, and $\bar{\mathbf{K}}$ is the corresponding kernel generated by the system matrices.

\subsubsection{Mamba and MambaIR}
Mamba \citep{gu2023mamba} extends this idea with a Selective State Space (S6) mechanism, where system parameters (such as $\bar{\mathbf{B}}, \mathbf{C}, \Delta$) are dynamically modulated by the input sequence. This allows the system to adaptively control signal propagation across time steps, enabling flexible modeling of long-range dependencies with linear computational cost.

MambaIR \citep{guo2024mambairv2, guo2025mambair} adapts the Mamba architecture for 2D image restoration, building upon the 2D Selective Scan (SS2D) module introduced in VMamba \citep{Liu2024vmamba}. The core strategy of SS2D is to model long-range dependencies by processing a flattened 1D representation of the image along multiple directional traversal paths.

Given an input feature map \(\mathbf{x} \in \mathbb{R}^{C \times H \times W}\), it is first flattened into a 1D sequence \(\mathbf{x}' \in \mathbb{R}^{(H \times W) \times C}\). This single sequence is then processed by four independent state-space models, each corresponding to a spatial direction \(\text{dir} \in \{\rightarrow, \leftarrow, \downarrow, \uparrow\}\).

For each direction, a directional operator \(\mathcal{F}_{\text{dir}}\) is applied to the shared input sequence to compute direction-specific features:
\begin{equation}
\mathbf{y}_{\text{dir}} = \mathcal{F}_{\text{dir}}(\mathbf{x}'),
\end{equation}
where \(\mathcal{F}_{\text{dir}}\) includes path-specific sequence reordering and structured recurrence via a 1D dynamic convolution kernel \(\bar{\mathbf{K}}_{\text{dir}}\), derived from the underlying SSM parameters.

The directional outputs are aggregated via summation and reshaped to restore the original spatial format:
\begin{equation}
\mathbf{y} = \text{Reshape}\left(\sum_{\text{dir}} \mathbf{y}_{\text{dir}}\right).
\end{equation}

To further stabilize training and preserve local information, MambaIR incorporates a Residual State-Space Block (RSSB), which includes two components: one is local enhancement, implemented via a convolutional layer to recover fine-grained details, and the other is channel attention, designed to reduce redundancy across channels caused by the large hidden dimensions typical in SSMs. In addition, MambaIR introduces a Residual State-Space Group (RSSG), which stacks multiple RSSBs with shared structural priors and inter-block residual connections. This hierarchical grouping allows MambaIR to model more expressive spatial features while maintaining stability and parameter efficiency.


\subsection{Overall Pipeline}

\begin{figure}[!t]
\centering
\includegraphics[width=0.9\textwidth]{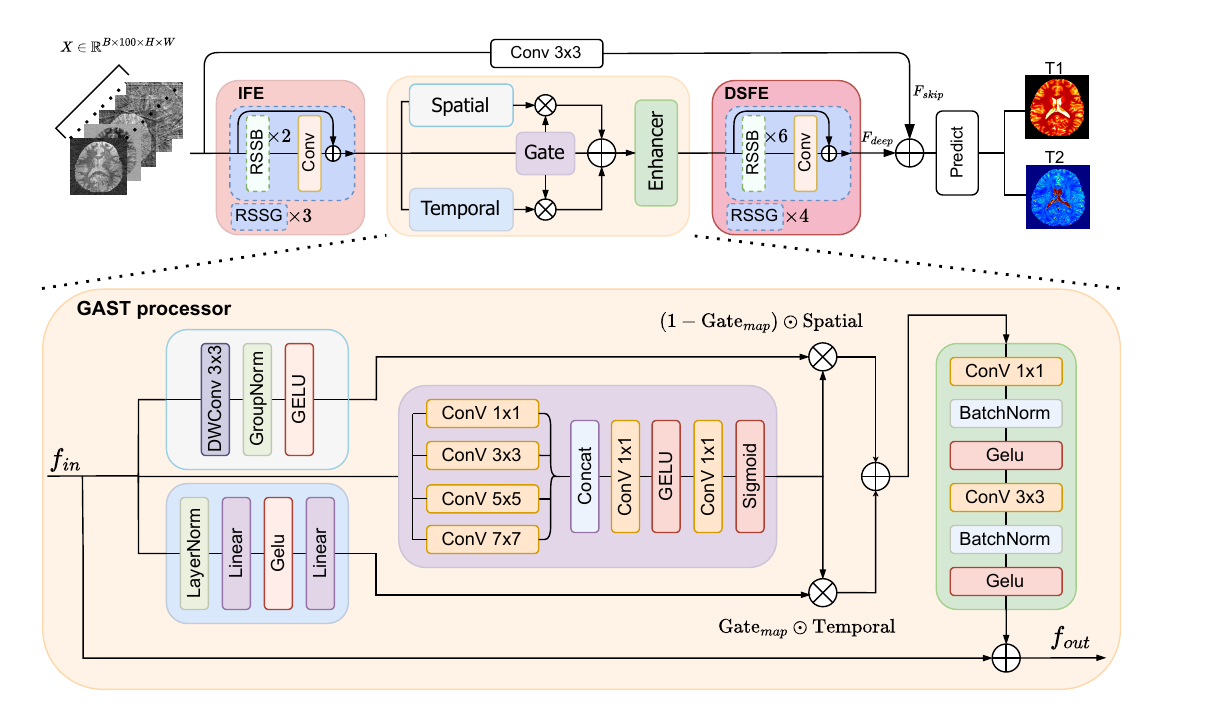}
   \caption{
Proposed GAST-Mamba network structure. The network consists of four major modules: Initial Feature Extractor (IFE), Gate-Aware Spatial-Temporal Adaptive Processor (GAST), Deep Spatial Feature Extractor (DSFE), and Prediction Head. Residual State-Space Block (RSSB) and Residual State-Space Block Groups (RSSG) are adopted from MambaIR.
}

\label{fig:GAST-mamba}
\end{figure}

The proposed GAST-Mamba network, illustrated in Figure~\ref{fig:GAST-mamba}, is an end-to-end framework that reconstructs T1 and T2 maps from input MRF signals. The network consists of four main components: 

1. \textbf{Initial Fingerprint Encoder (IFE):}  
The raw MRF input \(X \in \mathbb{R}^{B \times 100 \times H \times W}\) is constructed by concatenating the real and imaginary parts of the complex-valued MRF signal, each reduced to 50 dimensions via SVD along the channel axis. This input is then processed by a MambaIR-based encoder to produce latent spatial features \(F_{\text{latent}} \in \mathbb{R}^{B \times 64 \times H \times W}\).

2. \textbf{Gate-Aware Spatial-Temporal Adaptive (GAST) Processor:}  
The latent feature map is refined through stacked GAST blocks incorporating dual-path Spatial-Temporal processing and multi-scale gating, resulting in enhanced features \(F_{\text{GAST\_out}} \in \mathbb{R}^{B \times 64 \times H \times W}\).

3. \textbf{Deep Spatial Feature Extractor (DSFE):}  
A secondary MambaIR encoder captures complementary spatial dependencies, yielding deep features \(F_{\text{deep}} \in \mathbb{R}^{B \times 64 \times H \times W}\).

4. \textbf{Prediction Head:}  
The final quantitative prediction is obtained by blending initial fingerprint features with deep spatial features via a learnable residual connection. The fused representation is then projected by a \(1 \times 1\) convolution to generate the output parameter maps.

\subsubsection{Initial Fingerprint Encoder (IFE)}

IFE transforms 100-channel MRF fingerprints into 64-channel latent representations via a three-stage hierarchical MambaIR encoder: 
\[
\mathbb{R}^{B \times 100 \times H \times W} \rightarrow \mathbb{R}^{B \times 64 \times H \times W}.
\]

Each IFE module consists of three stacked residual state-space block group blocks, where each residual state-space block groups contain two residual state-space blocks  connected via residual shortcuts. The internal embedding dimension is set to 96, with a state size of 10 and a feedforward expansion ratio of 1.2. This hierarchical structure facilitates progressive spatial-temporal representation learning from compressed MRF fingerprints, while preserving gradient stability during training. The implementation details of the MambaIR-based RSSB follow the original formulation in~\citep{guo2025mambair}. 

\subsubsection{Gate-Aware Spatial-Temporal Adaptive Processor (GAST)}

\paragraph{Design Motivation}

In MRF reconstruction, the encoded fingerprint features exhibit strong spatial heterogeneity across anatomical regions, reflecting variations in tissue properties and structural complexity. This heterogeneity poses a significant challenge for conventional deep learning models, which must balance two competing objectives: preserving fine-grained local details (such as tissue boundaries and textures) and maintaining global spatial consistency across broader anatomical regions.

Spatial feature extractors (e.g., depthwise convolutions) are effective at capturing fine structural patterns but often lack contextual awareness, leading to spatial discontinuities. In contrast, temporal processing modules (e.g., MLPs or self-attention) model long-range dependencies but may oversmooth critical anatomical boundaries.

To address this trade-off, we propose the Gate-Aware Spatial-Temporal Adaptive (GAST) processor, which adaptively integrates spatial and temporal features via a spatially-varying gate mechanism. This design enables each spatial location to selectively emphasize spatial detail or temporal context, depending on the underlying anatomical structure.

\paragraph{Dual-Path Feature Extraction}

The GAST block implements parallel spatial and temporal feature processing pathways to address the complementary nature of spatial detail preservation and long-range dependency modeling.

Spatial Pathway employs depthwise separable convolutions to capture fine-grained spatial features while maintaining computational efficiency:

\begin{equation}
F_{\text{spatial}} = \text{GELU}(\text{GroupNorm}(\text{DWConv}_{3 \times 3}(F_{\text{latent}})))
\end{equation}

Here, the depthwise convolution with \( 3 \times 3 \) kernels captures spatially patterns in a channel-wise manner, enhancing fine-grained structural details. Group normalization with \( \max(C // 8,\ 1) \) groups provides stable training dynamics while preserving per-channel characteristics, followed by GELU activation for non-linear feature refinement.

Temporal Pathway performs channel-wise transformations to model long-range dependencies without spatial constraints. Given an input feature map \( F_{\text{latent}} \in \mathbb{R}^{B \times C \times H \times W} \), it is first reshaped into a sequence \( F \in \mathbb{R}^{BHW \times C} \), where each spatial position is treated as an independent token. The sequence is normalized using layer normalization, followed by a two-layer MLP with a \(4\times\) expansion ratio and GELU activation:
\begin{equation}
F_{\text{mlp}} = \text{Linear}(C \rightarrow 4C) \rightarrow \text{GELU} \rightarrow \text{Linear}(4C \rightarrow C).
\end{equation}

Finally, the output is reshaped back to the original spatial format \( F_{\text{temporal}} \in \mathbb{R}^{B \times C \times H \times W} \). This design enables dense inter-channel interaction while preserving the spatial layout. Layer normalization stabilizes training during temporal feature mixing.




\paragraph{Multi-Scale Gate Network}







To enhance spatial feature representation, we propose a multi-scale gate network (as shown in Fig.~\ref{fig:GAST-mamba}), which consists of parallel depthwise convolutional branches with varying kernel sizes (e.g., \(1 \times 1\), \(3 \times 3\), \(5 \times 5\), and \(7 \times 7\)). Each branch captures features at a different spatial scale, and their outputs are concatenated to form a unified multi-scale representation, denoted as \(G_{\text{concat}}\).

The concatenated features are subsequently fused through a spatially adaptive gating mechanism:
\begin{equation}
\text{Gate}_{\text{map}} = \sigma\left( \text{Conv}_{1 \times 1} \left( \text{GELU} \left( \text{Conv}_{1 \times 1} \left( G_{\text{concat}} \right) \right) \right) \right)
\end{equation}

where \( \sigma(\cdot) \) denotes the Sigmoid function. The learned gate map provides spatially-adaptive weights that determine the optimal balance between spatial and temporal processing at each pixel location.

The spatially adaptive fusion is formally defined as:

\begin{equation}
\label{eq:gate-map}
F_{\text{output}} = \text{Gate}_{\text{map}} \odot F_{\text{spatial}} + (1 - \text{Gate}_{\text{map}}) \odot F_{\text{temporal}},
\end{equation}

where \( \odot \) denotes element-wise multiplication. The gate map is predicted from the multi-scale feature fusion and encodes the blending ratio between spatial and temporal pathways for each spatial location. Values closer to 1 emphasize spatial consistency, while values near 0 highlight temporal details. This mechanism enables content-aware spatial adaptation to heterogeneous tissue properties in MRF data.

\paragraph{Feature Refinement}
The adaptively fused features are further refined through a lightweight enhancement block:
\begin{equation}
F_{\text{enhanced}} = \text{Conv}_{1 \times 1}\left( \text{GELU}\left( \text{GroupNorm}\left( \text{DWConv}_{3 \times 3}(F_{\text{output}}) \right) \right) \right)
\end{equation}

Finally, the refined output is combined with the original input via a learnable residual connection:
\begin{equation}
F_{\text{GAST\_out}} = \alpha \cdot F_{\text{enhanced}} + F_{\text{latent}},
\end{equation}
where \( \alpha \in \mathbb{R} \) is a learnable scalar initialized to 1.0, and \( F_{\text{latent}} \) denotes the original input to the GAST module. This design facilitates gradient flow and provides additional modeling capacity for feature enhancement.



\subsubsection{Deep Spatial Feature Extractor (DSFE)}

The DSFE module refines the 64-channel feature map output from the GAST processor using a deeper four-stage MambaIR encoder:
\[
\mathbb{R}^{B \times 64 \times H \times W} \rightarrow \mathbb{R}^{B \times 64 \times H \times W}.
\]
It consists of four stacked RSSG blocks, each containing six RSSBs with residual connections. The embedding dimension is set to 64, with a state size of 10 and a feedforward expansion ratio of 1.2. Compared to the shallower IFE module, the increased depth of DSFE enables more expressive modeling of anatomical structures and long-range spatial dependencies, which is critical for accurate quantitative mapping of T1 and T2 values.

\subsubsection{Prediction Head}

The final quantitative estimation employs a learnable feature blending strategy to combine 
initial fingerprint features and deep spatial representations:

\begin{equation}
F_{\text{blended}} = \beta \cdot F_{\text{skip}} + (1 - \beta) \cdot F_{\text{deep}},
\end{equation}

where \( F_{\text{skip}} = \text{Conv}_{3 \times 3}(X) \in \mathbb{R}^{B \times 64 \times H \times W} \) 
represents the initially processed MRF fingerprints, 
\( F_{\text{deep}} \in \mathbb{R}^{B \times 64 \times H \times W} \) denotes the output from DSFE, 
and \( \beta \in \mathbb{R} \) is a learnable scalar initialized to 0.5. 
This blending mechanism allows the model to adaptively balance low-level fingerprint information 
with high-level spatial features.

Subsequently, a single \(1 \times 1\) convolution maps the blended features to the target quantitative parameter space:

\begin{equation}
\hat{Y} = \text{Conv}_{1 \times 1}(F_{\text{blended}}) \in \mathbb{R}^{B \times 2 \times H \times W},
\end{equation}

producing the predicted T1 and T2 relaxation maps. This design integrates spatial detail preservation with 
robust high-level modeling, supporting accurate and reliable quantitative estimation.

\subsection{Data Generation and Preprocessing}
\label{sec:generate-data}

\subsubsection{Simulation Dataset}

\begin{figure}[!t]
\centering
\includegraphics[width=0.9\textwidth]{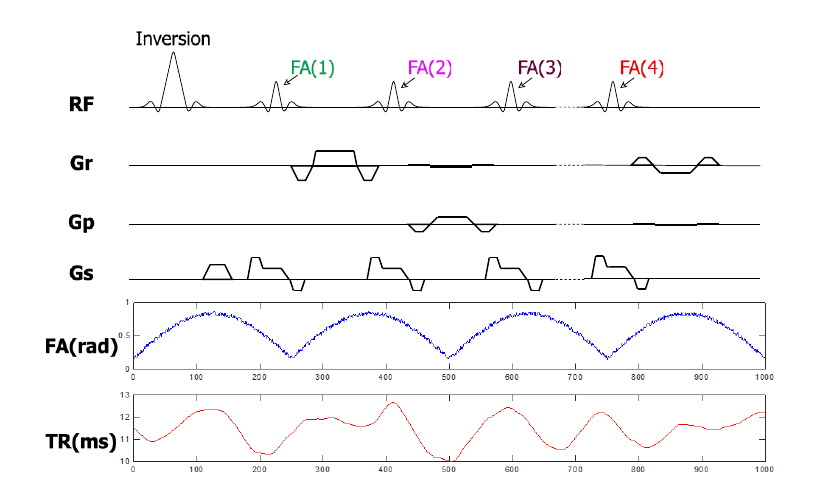}
   \caption{Inversion-recovery balanced steady-state free precession (IR-bSSFP) sequence, FAs, and TRs applied in this work}
\label{fig:sequence}
\end{figure}

\begin{figure}[!t]
\centering
\includegraphics[width=0.9\textwidth]{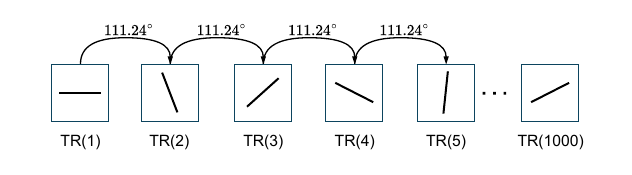}
   \caption{Radial trajectory applied in this work.}
\label{fig:radial-arm}
\end{figure}

Supervised training data were generated by simulating single-shot radial MRF acquisitions using the inversion-recovery balanced steady-state free precession (IR-bSSFP) sequence. This sequence comprises 1000 repetitions with variable flip angles (FAs) and repetition times (TRs). Figure~\ref{fig:sequence} and Figure~\ref{fig:radial-arm} demonstrates the sequence setting and single shot radial arm trajectory applied in this work. Given voxel-wise tissue parameters \(T_1\), \(T_2\), and off-resonance frequency \(B_0\), the temporal signal evolution was simulated using the Bloch equation:
\begin{equation}
S_{\text{image}} = \text{Bloch}(T_1, T_2, B_0, \text{FAs}, \text{TRs}).
\end{equation}

To emulate accelerated acquisition, the 1000-point time series was truncated to 200 frames prior to k-space simulation. We implemented a radial undersampling scheme following the method described in \citep{cloos2019rapid}, using golden-angle rotated spokes at each time frame. The resulting undersampled k-space was computed via non-uniform FFT (NUFFT):
\begin{equation}
S_{\text{kspace}} = \text{NUFFT}(S_{\text{image}}, k_{\text{traj}}),
\end{equation}
where \(k_{\text{traj}}\) denotes the rotated radial trajectory. The corresponding aliased image series was reconstructed using inverse NUFFT:
\begin{equation}
S_{\text{alias}} = \text{NUFFT}^{-1}(S_{\text{kspace}}, k_{\text{traj}}).
\end{equation}

Ground-truth parameter maps of \(T_1\), \(T_2\)\footnote{T2\textsuperscript{*} maps were acquired from ME-MP2RAGE scans, and T2 values were estimated by scaling T2\textsuperscript{*} with a factor of 1.5.}, and \(B_0\) were derived from ME-MP2RAGE datasets \citep{sun2020extracting}. Each simulated MRF sample had a spatial resolution of \(192 \times 192\) and 200 temporal frames after truncation.

To reduce memory requirements and preserve essential signal characteristics, singular value decomposition (SVD) \citep{mcgivney2014svd} was used to compress each complex-valued time series from 200 to 50 components. The top 50 singular vectors were derived from the MRF dictionary and used to project the real and imaginary parts separately, yielding an input volume of shape \(192 \times 192 \times 100\). The corresponding T1 and T2 maps were used as regression targets during training.

To investigate the effect of acquisition length, additional datasets were generated by truncating the time series to \(t = 400, 600, 800,\) and \(1000\) frames prior to SVD. The \(t = 200\) configuration used in all main experiments corresponds to a five-fold (5 $\times$) acceleration over the full-length sequence, enabling substantial scan time reduction.

\subsubsection{\textit{In Vivo} Dataset}

\textit{In vivo} data were acquired from a healthy volunteer on a 3T Siemens Prisma scanner (Erlangen, Germany) using a 32-channel head coil. A total of 32 axial slices were scanned using a 1000-timepoint IR-bSSFP sequence, with an in-plane resolution of 1 mm and slice thickness of 3 mm. The pulse sequence, including FAs, TRs, and inversion recovery timing, was identical to that used in simulation, and golden-angle radial sampling was applied at each frame. Written informed consent was obtained prior to acquisition.

To align with the training data format, the acquired MRF signals were truncated to 200 time points and projected onto the same SVD basis used in the simulation preprocessing, yielding an input size of \(192 \times 192 \times 100\). This dataset was excluded from all training and validation procedures and was used solely to evaluate the model's generalizability to \textit{in vivo} acquisitions.


\section{Experiments}
\label{sec:experiments}

We refer to Sec.~\ref{sec:generate-data} for details on data simulation. This section describes the experimental setup, including training configurations and evaluation metrics, followed by quantitative and qualitative comparisons on simulated and in vivo datasets. We also present ablation studies to assess the contribution of individual architectural components.

\subsection{Experimental Settings}
\label{sec:experimental-settings}

\paragraph{Training Configuration}
The model was trained on 1300 simulated samples, each with spatial dimensions \(192 \times 192\) and 100 temporal channels. The dataset was randomly partitioned into 1170 training and 130 validation samples using a fixed random seed to ensure reproducibility. A separate test set of 90 samples from two unseen simulated objects was used for evaluation.

Training was performed using the AdamW optimizer with a batch size of 2 for 100 epochs. A learning rate of \(5 \times 10^{-5}\) was applied uniformly to all modules, with a weight decay of 0.01. A MultiStepLR scheduler was used to reduce the learning rate by a factor of 0.5 at epochs 25, 50, 75, and 90.

The total loss \(\mathcal{L}_{\text{total}}\) combined MSE and \(L_1\) terms for T1 and T2, with a dynamic T1-specific weight \(w(e)\) that linearly decayed from 1.5 to 1.0 over epochs \(e\):
\begin{align}
\mathcal{L}_{\text{total}}^{(e)} =\;& 
w(e) \cdot \left[ \mathcal{L}_{\text{MSE}}(T_1, \hat{T}_1) 
+ \alpha \cdot \mathcal{L}_{1}(T_1, \hat{T}_1) \right] \notag \\
& + \left[ \mathcal{L}_{\text{MSE}}(T_2, \hat{T}_2) 
+ \alpha \cdot \mathcal{L}_{1}(T_2, \hat{T}_2) \right],
\end{align}
where \(\alpha = 0.2\) and \(\hat{T}_1, \hat{T}_2\) denote the predicted outputs.

Ground truth T1 and T2 maps were standardized using global mean and standard deviation values computed from all non-background voxels in the training set. All models were trained on an NVIDIA RTX 5090 GPU, requiring approximately 8 hours.

\paragraph{Evaluation Metrics}
We used four standard metrics to assess reconstruction performance: peak signal-to-noise ratio (PSNR), structural similarity index (SSIM), root mean square error (RMSE), and normalized mean squared error (NMSE). These metrics were computed separately for T1 and T2 over non-background voxels. Final results were reported as the mean ± standard deviation across all test slices.

\paragraph{Comparative Methods}
\label{para:compare-method}
To evaluate the effectiveness of our proposed GAST-Mamba, we compared it against four representative deep learning models for MRF reconstruction: SCQ~\citep{fang2019deep}, LG-ViT~\citep{li2024deep}, CONV-ICA~\citep{soyak2021channel}, and MRF-Mixer~\citep{ding2025mrf}. These methods represent diverse architectural paradigms in MRF reconstruction, each contributing unique strategies for modeling spatial-temporal features.

For fair comparison, all models were trained on the same simulated MRF dataset with matched preprocessing, loss functions, and input dimensionality. Where applicable, we reproduced results using publicly available code and default training configurations. For SCQ and LG-ViT, we followed their released implementations and training strategies, making minimal modifications to adapt the input format and normalization to our simulated MRF data.  CONV-ICA was reimplemented based on the architecture and training protocol described in the original publication. MRF-Mixer, based on our previous work, was adapted to accept the new input format while preserving the core design.

\subsection{Simulation-Based Results}
\input{tables/table1}

\begin{figure}[!t]
\centering
\includegraphics[width=0.9\textwidth]{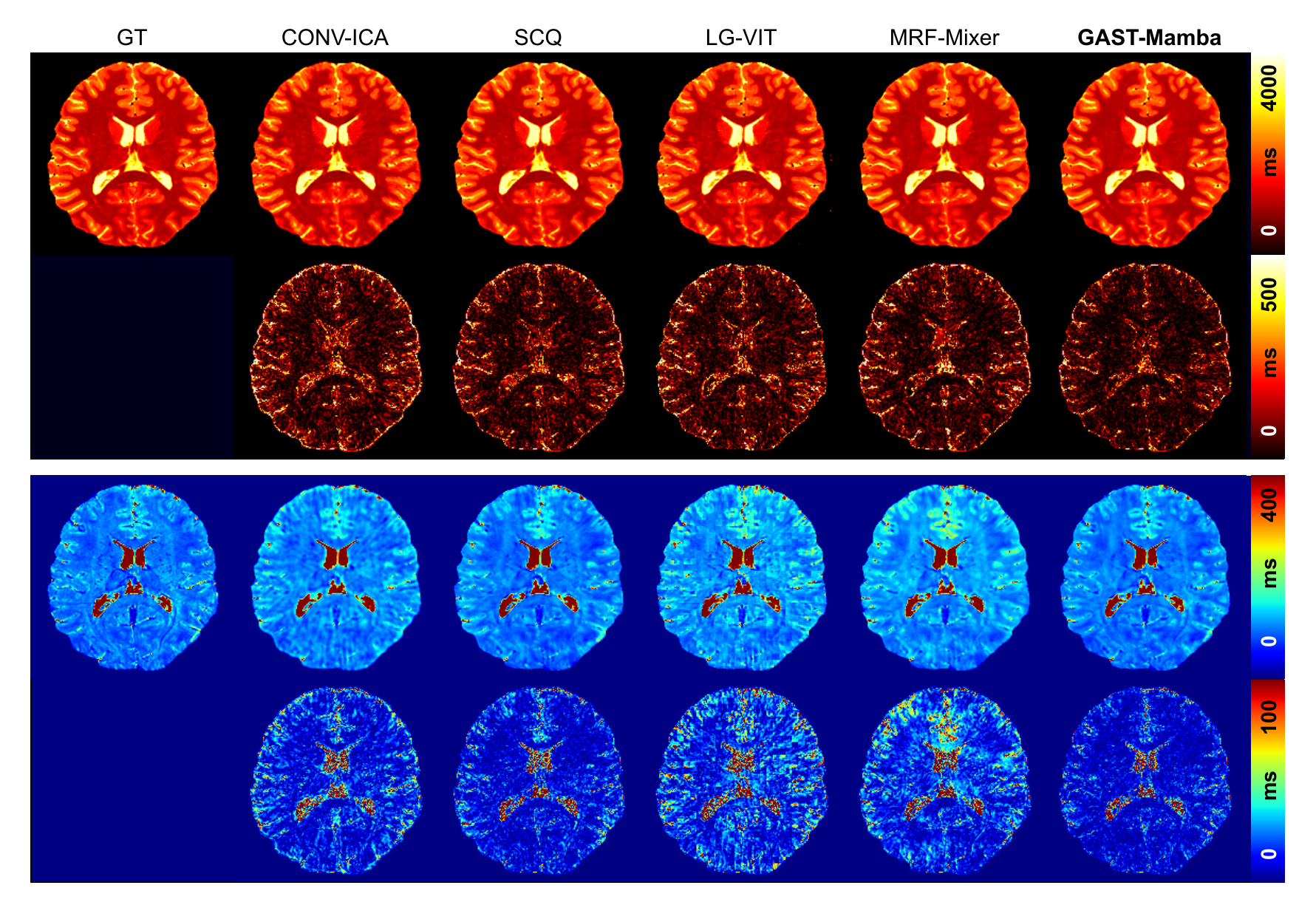}
   \caption{
Qualitative comparison of T1 and T2 map reconstructions on 5 $\times$ accerlated simulated data with known ground truth (GT). Each row corresponds to a different T1/T2 combination, and columns show the outputs of different models, including CONV-ICA, SCQ, LG-ViT, MRF-Mixer, and the proposed GAST-Mamba. For each sample, the reconstructed map and corresponding absolute error map are shown.
}

\label{fig:simulation-1}
\end{figure}

Table~\ref{tab:t1t2-results} summarizes the quantitative performance of different models on the T1 and T2 map reconstruction tasks. Evaluation metrics include PSNR and SSIM, reported as mean~$\pm$~standard deviation over the test set. The proposed GAST-Mamba model achieves the best overall performance, attaining the highest PSNR and SSIM values for both T1 and T2 reconstructions. Specifically, it achieves 33.12~dB PSNR and 0.9674 SSIM for T1, and 30.62~dB PSNR and 0.9124 SSIM for T2.

Figure~\ref{fig:simulation-1} shows representative visual comparisons on 5 times accerlated simulated data. GAST-Mamba yields reconstructed T1 and T2 maps that are visually closest to the ground truth, with minimal residuals and clearly defined structural boundaries.

In the T2 maps, LG-ViT and CONV-ICA exhibit structured errors predominantly within the white matter regions, suggesting inaccurate signal recovery in homogeneous tissues. MRF-Mixer shows high residuals in a partial volume region near the superior frontal cortex, where gray and white matter signals are mixed, resulting in structured aliasing artifacts along the interhemispheric fissure. SCQ achieves better contrast than LG-ViT and MRF-Mixer but still displays elevated residuals across the entire map, with overall higher error magnitudes compared to GAST-Mamba. These observations are consistent with the quantitative trends reported in Table~\ref{tab:t1t2-results}

\begin{figure}[!t]
\centering
\includegraphics[width=1.0\linewidth]{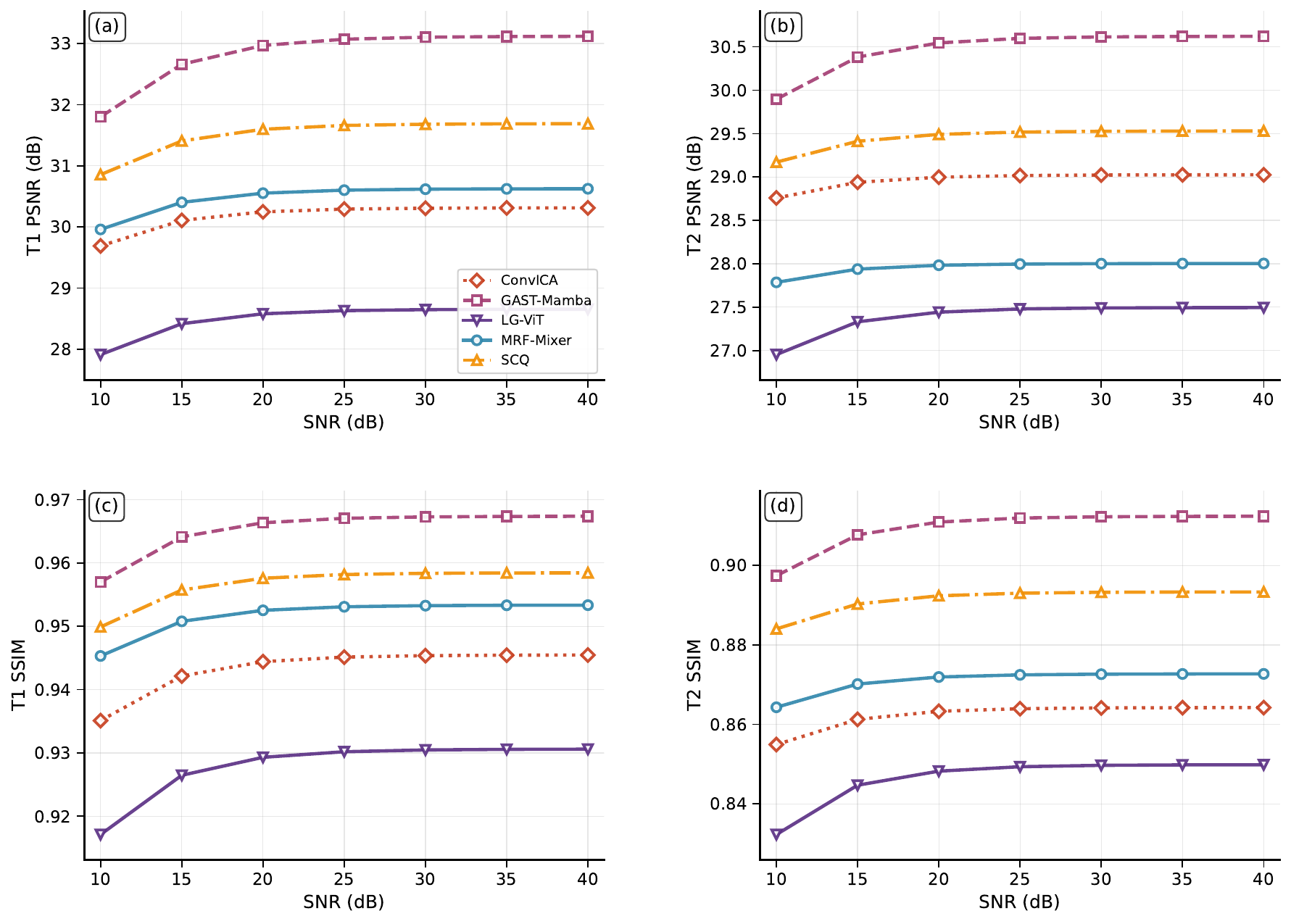}
   \caption{
Quantitative evaluation of model robustness to noise. PSNR and SSIM metrics for T1 (a, c) and T2 (b, d) map reconstruction are plotted across varying SNR levels (10--40~dB).
}

\label{fig:simulation-snr}
\end{figure}

Figure~\ref{fig:simulation-snr} presents the reconstruction performance of all models under varying signal-to-noise ratio (SNR) levels from 10~dB to 40~dB. GAST-Mamba consistently outperforms all baselines across the entire noise range, achieving the highest PSNR and SSIM values on both T1 and T2 maps. 

\begin{figure}[!t]
\centering
\includegraphics[width=1.0\linewidth]{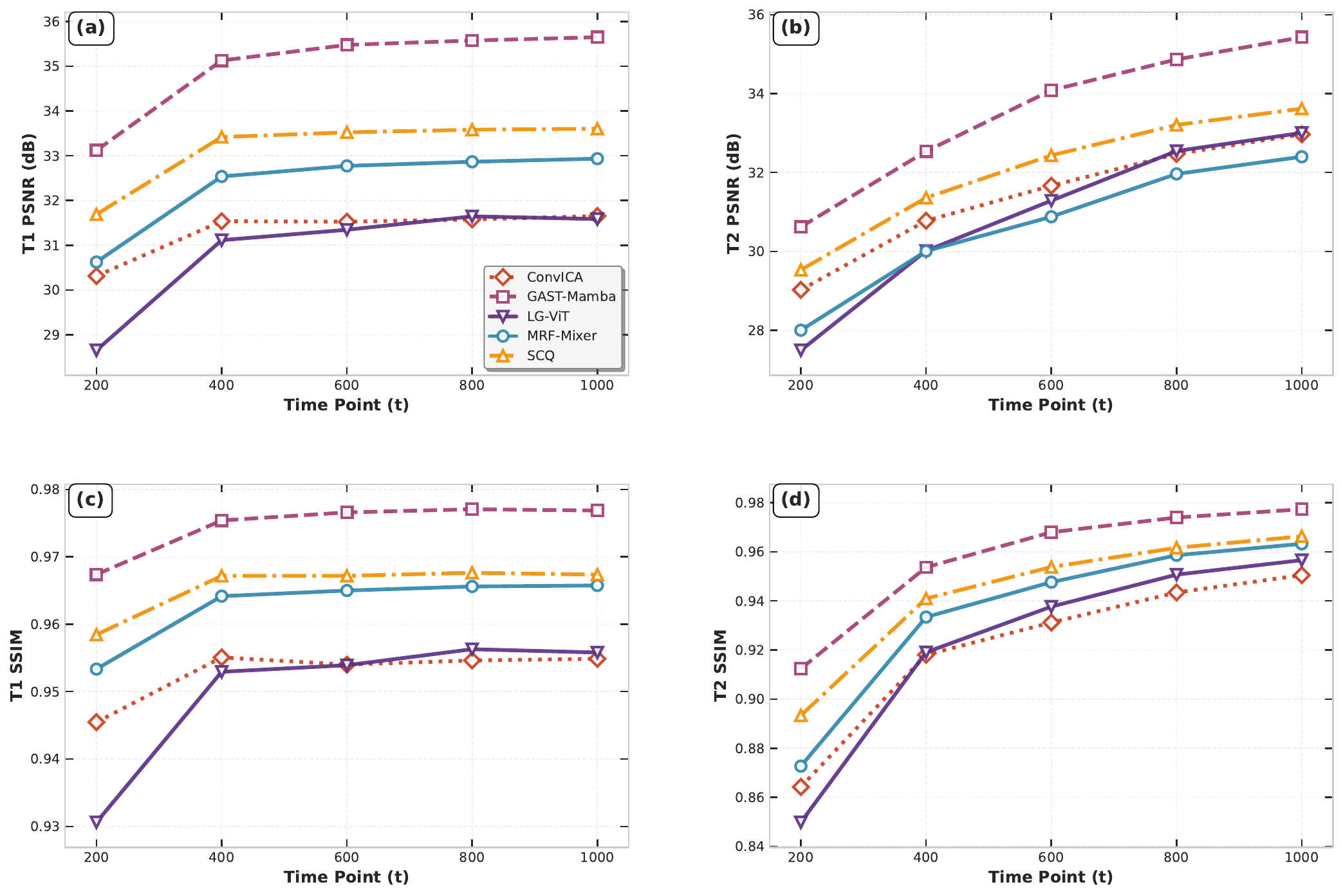}
   \caption{
Quantitative evaluation of model performance under different acquisition lengths. PSNR and SSIM metrics for T1 (a, c) and T2 (b, d) map reconstruction are plotted for time points ranging from \( t=200\) to \( t=1000\).}

\label{fig:simulation-differ-t}
\end{figure}

\input{tables/table-different-t-compare}

Figure~\ref{fig:simulation-differ-t} shows the PSNR and SSIM performance of all models from \(t = 200\) to \(t = 1000\), where GAST-Mamba consistently achieves the highest values on both T1 and T2 maps. Table~\ref{tab:model-different-t} presents the corresponding RMSE and NMSE results, further confirming that GAST-Mamba produces the lowest reconstruction errors across all time points.

\subsection{\textit{In Vivo} Evaluation}

\begin{figure}[!t]
\centering
\includegraphics[width=0.95\textwidth]{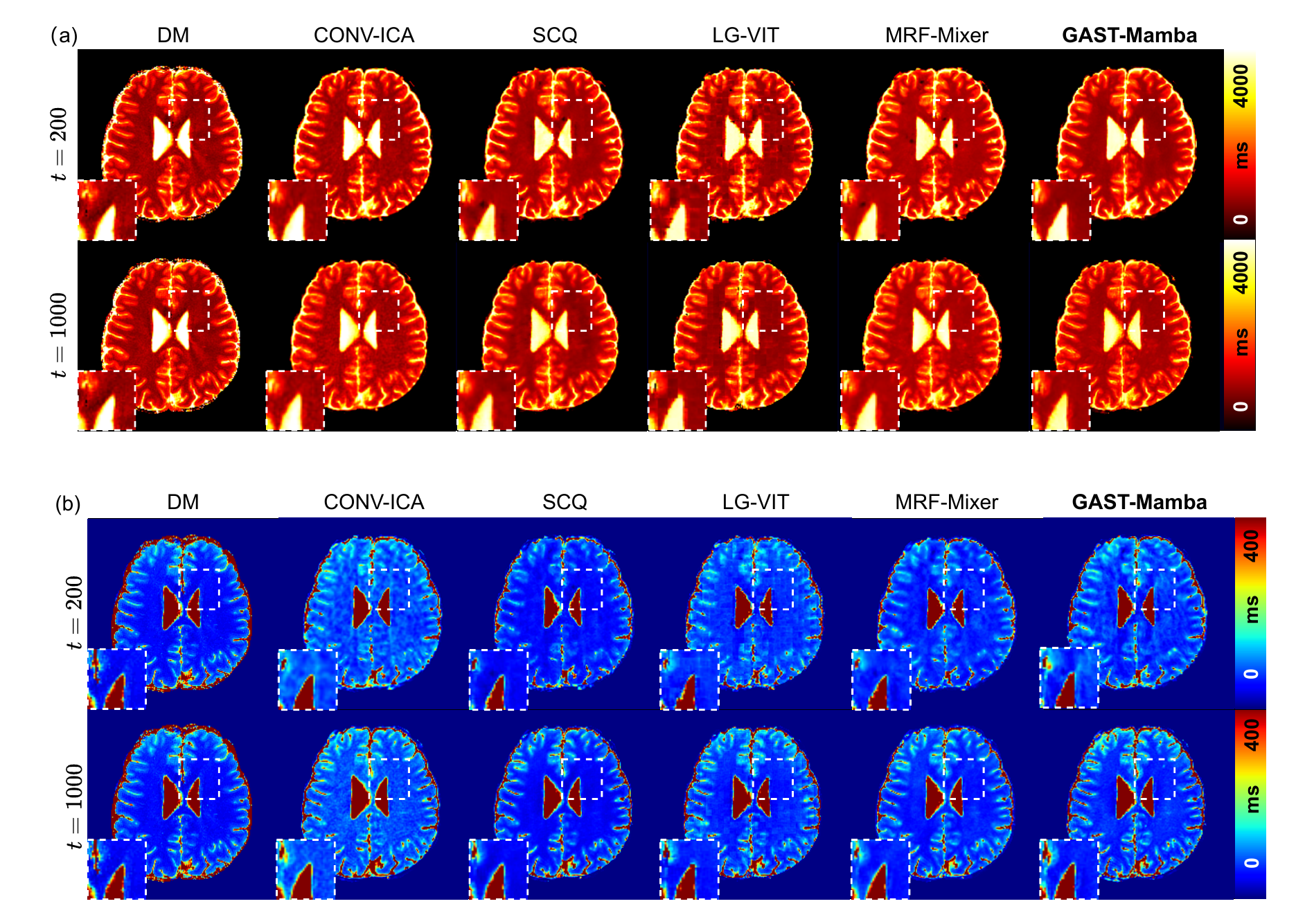}
\caption{
Qualitative comparison of T1 and T2 map reconstructions on in vivo brain data. Columns represent different reconstruction methods, including dictionary matching (DM), CONV-ICA, SCQ, LG-ViT, MRF-Mixer, and the proposed GAST-Mamba. (a) T1 reconstruction results for $t=200$ and $t=1000$, (b) T2 reconstruction results for $t=200$ and $t=1000$,
}
\label{fig:invivo-1}
\end{figure}

Figure~\ref{fig:invivo-1} shows the reconstruction results on 5$\times$ accelerated \textit{in vivo} MRF data. Overall, GAST-Mamba produces visually sharper T1 and T2 maps with improved anatomical consistency compared to all baseline methods. Cortical and subcortical structures are clearly delineated, and tissue boundaries remain smooth and continuous without introducing visible artifacts.

For T1 reconstruction, CONV-ICA, SCQ, and MRF-Mixer achieve globally coherent maps that are qualitatively comparable to GAST-Mamba in terms of large-scale structure. However, GAST-Mamba yields cleaner results with minimal aliasing artifacts. In contrast, CONV-ICA presents visible aliasing patterns, particularly in white matter regions. SCQ and MRF-Mixer exhibit localized information loss in selected regions, where structural details such as sulci or tissue interfaces are partially obscured. LG-ViT demonstrates severe spatial artifacts across the entire map, while DM, constrained by the single-shot acquisition and lack of learning-based regularization, shows strong noise and reduced contrast.

For T2 reconstruction, only GAST-Mamba, SCQ, and MRF-Mixer are able to produce structurally plausible maps. Among these, GAST-Mamba recovers finer structural detail, especially at tissue boundaries, where signal transitions are more accurately preserved. SCQ and MRF-Mixer retain coarse structural information but exhibit moderate smoothing and detail loss. In contrast, CONV-ICA and LG-ViT fail to reconstruct key anatomical features and display prominent degradation, limiting their utility for reliable T2 quantification.

\input{tables/table_ablation}
\subsection{Ablation Study}

\input{tables/table-ablation-results}

To validate the effectiveness of each component in our proposed GAST-Mamba architecture, we conducted ablation experiments by systematically removing key modules (Table~\ref{tab:ablation_configurations}): (A1) without the Deep Spatial Feature Extractor (DSFE), (A2) without the GAST processor, (A3) using only the Initial Fingerprint Encoder (IFE), and (A4) replacing IFE with a CNN encoder. The results in Table~\ref{tab:ablation-study} show that the complete model achieves the highest or comparable performance across all timepoints, consistently ranking among the top-performing configurations for both T1 and T2 reconstruction. Most notably, the GAST processor provides significant benefits under highly undersampled conditions—at $t=200$, our full model achieved 33.12~$\pm$~1.09~dB PSNR compared to 32.58~$\pm$~1.10~dB for A1, 33.00~$\pm$1.10~dB for A2, 32.63~$\pm$~1.10~dB for A3, and 32.70$\pm$1.09~dB for A4, reflecting substantial improvements under highly undersampled acquisition conditions.

The importance of GAST becomes particularly evident in challenging acquisition scenarios, where adaptive spatial-channel processing is critical. While A2 (without GAST) performed comparably at longer timepoints, it showed consistent degradation at earlier timepoints, where the gate-aware mechanism's ability to emphasize informative features and suppress noise becomes more valuable. Likewise, the Mamba-based encoder in the full model outperformed the CNN-based variant (A4), particularly under high acceleration, demonstrating the advantages of state-space modeling in capturing temporal dynamics from undersampled MRF sequences. These results confirm that each component contributes meaningfully to the overall reconstruction quality, with GAST offering essential adaptive processing capabilities for robust performance under clinically relevant acceleration factors.


\section{Discussion}

\subsection{Key Findings and Implications}

Our results demonstrate that GAST-Mamba achieves substantial improvements in MRF reconstruction quality, particularly under highly accelerated acquisition conditions. The most significant finding is the model's robust performance at t200, where severe undersampling typically degrades reconstruction quality. At this challenging timepoint, GAST-Mamba achieved 33.12$\pm$1.09 dB PSNR for T1 reconstruction, representing a meaningful improvement over ablation variants that lack key architectural components. This performance gain is clinically significant, as it enables reliable parameter estimation from acquisitions that would otherwise be considered too accelerated for accurate tissue characterization.

The ablation study reveals that GAST's adaptive channel processing becomes increasingly critical as acquisition acceleration increases. The gate-aware mechanism's ability to selectively emphasize informative features while suppressing noise-dominated channels proves essential when working with limited MRF timepoints. This finding has important implications for understanding how deep learning models can be specifically designed to handle the unique challenges of accelerated MRI, where traditional reconstruction methods often fail due to insufficient data constraints.

\subsection{Architectural Innovation and Computational Efficiency}

The dual Mamba architecture addresses fundamental limitations of existing approaches through complementary mechanisms. While CNN-based methods like SCQ and MRF-Mixer are constrained by limited receptive fields, and transformer-based approaches such as LG-ViT suffer from quadratic computational complexity, our Mamba-based design achieves global spatial modeling with linear complexity. This efficiency gain is particularly important for high-resolution MRF data, where memory constraints often limit the applicability of attention-based models.

The hierarchical processing pipeline—from initial fingerprint encoding through adaptive channel processing to deep spatial feature extraction—enables multi-scale feature learning that captures both spatial tissue heterogeneity and temporal anatomical context. This design differs from existing methods that typically rely on single-scale processing, explaining the consistent performance improvements observed across different acceleration factors and tissue types.

\subsection{Clinical Translation and Practical Impact}

The demonstrated ability to reconstruct high-quality T1 and T2 maps from 5$\times$ accelerated single-shot acquisition senables clinical translation in time-sensitive scenarios. Conventional single-slice MRF acquisitions require 10-30 seconds per slice \citep{ma2013magnetic,ma2018fast,liu2023assessment}, while our approach maintains reconstruction fidelity at significantly reduced acquisition times. This acceleration capability is particularly valuable in pediatric imaging, patient populations with limited breath-holding capacity, and high-throughput clinical workflows where scan time directly impacts patient throughput.

Furthermore, the model's robustness across varying SNR conditions suggests potential applicability beyond conventional 3T systems. In particular, low-field and portable MRI systems—where limited signal strength often compromises quantitative imaging—could benefit from the enhanced noise resilience demonstrated by GAST-Mamba. This capability may help expand access to quantitative MRI in resource-limited settings or point-of-care applications.

\subsection{Limitations and Future Directions}

While the proposed method demonstrates strong reconstruction performance on both simulated and \textit{in vivo} data, several limitations remain. The \textit{in vivo} evaluation was performed on a single healthy subject, which does not capture the full variability in anatomy, scanner configurations, or acquisition protocols encountered in clinical practice. In addition, the model has not been tested on pathological cases, making its diagnostic reliability in the presence of lesions or structural abnormalities uncertain. Future validation on larger, multi-site, and pathology-rich datasets is necessary to assess both generalizability and clinical relevance.

Moreover, the current implementation is limited to 2D reconstruction on individual slices, which may introduce inter-slice inconsistencies and fail to fully capture the spatial continuity of volumetric brain structures. Extending the framework to 3D reconstruction would enable more anatomically coherent parameter maps and improve performance in applications requiring volumetric consistency, such as lesion quantification or longitudinal studies.

To further improve reconstruction accuracy, one promising direction is to incorporate partial volume (PV) information as a structural prior. Many voxels, especially near tissue boundaries, contain a mixture of tissue types, introducing ambiguity in signal interpretation. Estimating sub-voxel tissue fractions and using them as auxiliary inputs or regularization constraints may help disambiguate signal components and enhance parameter estimation fidelity.

Another potential direction involves unsupervised or self-supervised learning to mitigate the reliance on simulation-based ground truth. Supervised models trained on synthetic data may not fully generalize to diverse clinical acquisitions. Leveraging intrinsic signal redundancy, physics-informed consistency, or data-driven priors may enable learning directly from \textit{in vivo} data, facilitating better domain adaptation and improving robustness in real-world deployment.

\section{Conclusion}

In this work, we proposed GAST-Mamba, a novel end-to-end deep learning framework for accurate and efficient multi-parametric map reconstruction in MRF. The proposed model integrates dual Mamba-based encoders with a GAST processor in a hierarchical architecture. The framework employs IFE for primary feature extraction from MRF signal evolutions, followed by GAST for adaptive channel processing, and DSFE for enhanced spatial modeling. Unlike conventional dictionary-based approaches or patch-based CNNs, GAST-Mamba jointly leverages spatial continuity and temporal fingerprint dynamics in a unified architecture. Through extensive experiments on simulated and in vivo datasets, GAST-Mamba demonstrated superior reconstruction accuracy and robustness compared to existing methods, particularly in challenging conditions such as low SNR and high undersampling. These results suggest that GAST-Mamba provides an effective and scalable solution for quantitative MRI reconstruction.





\end{document}

%% file: tables/table1.tex

\begin{table*}[htbp]
\centering
\caption{Quantitative results (mean $\pm$ std) on T1 and T2 maps across different models. Best results are highlighted in \textbf{bold}.}
\begin{adjustbox}{width=\textwidth}
\begin{tabular}{lcccc}
\hline
\multirow{2}{*}{\textbf{Model}} & \multicolumn{2}{c}{\textbf{T1}} & \multicolumn{2}{c}{\textbf{T2}} \\
\cmidrule(lr){2-5} 
& PSNR & SSIM &  PSNR & SSIM \\
\hline
\textbf{GAST-Mamba} & \textbf{33.12 $\pm$ 1.09} & \textbf{0.9674 $\pm$ 0.0070} & \textbf{30.62 $\pm$ 1.54} & \textbf{0.9124 $\pm$ 0.0138}  \\
SCQ                 & 31.69 $\pm$ 1.16 & 0.9584 $\pm$ 0.0088 &  29.53 $\pm$ 1.54 & 0.8934 $\pm$ 0.0172  \\
LGVIT               & 28.65 $\pm$ 2.34 & 0.9306 $\pm$ 0.0413 &  27.50 $\pm$ 1.17 & 0.8499 $\pm$ 0.0310  \\
CONV-ICA            & 30.31 $\pm$ 0.99 & 0.9455 $\pm$ 0.0117 &  30.06 $\pm$ 4.43 & 0.8642 $\pm$ 0.0191  \\
MRF-Mixer           & 30.62 $\pm$ 1.12 & 0.9533 $\pm$ 0.0097 &  28.00 $\pm$ 1.14 & 0.8727 $\pm$ 0.0174  \\
\hline
\end{tabular}
\end{adjustbox}
\label{tab:t1t2-results}
\end{table*}

%% file: tables/table-different-t-compare.tex
\begin{table}[htbp]
\centering
\caption{Quantitative results (mean $\pm$ std) on T1 and T2 maps across different models. Best results for each timepoint are highlighted in bold.}
\label{tab:model-different-t}
\resizebox{\textwidth}{!}{%
\begin{tabular}{ll|ccccc}
\toprule
\multirow{2}{*}{\textbf{Metric}} & \multirow{2}{*}{\textbf{Timepoint}} & \multicolumn{5}{c}{\textbf{Method}} \\
\cmidrule(lr){3-7}
& & \textbf{GAST-Mamba} & \textbf{MRF-Mixer} & \textbf{SCQ} & \textbf{CONV-ICA} & \textbf{LG-ViT} \\
\midrule
\multicolumn{7}{l}{\textit{RMSE (T1)}} \\
& t200 & \textbf{85.67} $\pm$ 10.03 & 145.43 $\pm$ 15.10 & 101.10 $\pm$ 12.69 & 118.20 $\pm$ 12.76 & 149.26 $\pm$ 69.31 \\
& t400 & \textbf{68.06} $\pm$ 8.53 & 116.83 $\pm$ 13.49 & 82.93 $\pm$ 11.16 & 102.76 $\pm$ 12.09 & 111.53 $\pm$ 50.23 \\
& t600 & \textbf{65.36} $\pm$ 8.10 & 113.69 $\pm$ 13.36 & 81.92 $\pm$ 11.08 & 102.91 $\pm$ 12.44 & 106.99 $\pm$ 34.79 \\
& t800 & \textbf{64.62} $\pm$ 7.90 & 112.45 $\pm$ 12.74 & 81.34 $\pm$ 10.86 & 102.29 $\pm$ 12.27 & 102.55 $\pm$ 25.27 \\
& t1000 & \textbf{64.06} $\pm$ 7.72 & 111.50 $\pm$ 12.22 & 81.14 $\pm$ 10.79 & 101.35 $\pm$ 11.91 & 103.16 $\pm$ 23.90 \\
\cmidrule(lr){2-7}
\multicolumn{7}{l}{\textit{RMSE (T2)}} \\
& t200 & \textbf{25.14} $\pm$ 4.31 & 42.98 $\pm$ 5.22 & 28.50 $\pm$ 4.92 & 30.06 $\pm$ 4.43 & 35.77 $\pm$ 4.99 \\
& t400 & \textbf{20.22} $\pm$ 3.73 & 34.28 $\pm$ 5.17 & 23.12 $\pm$ 4.10 & 24.60 $\pm$ 3.76 & 26.82 $\pm$ 4.19 \\
& t600 & \textbf{16.94} $\pm$ 3.16 & 30.96 $\pm$ 4.37 & 20.43 $\pm$ 3.62 & 22.23 $\pm$ 3.49 & 23.24 $\pm$ 3.95 \\
& t800 & \textbf{15.48} $\pm$ 2.91 & 27.43 $\pm$ 4.50 & 18.70 $\pm$ 3.32 & 20.28 $\pm$ 3.31 & 20.12 $\pm$ 3.45 \\
& t1000 & \textbf{14.52} $\pm$ 2.79 & 26.08 $\pm$ 4.20 & 17.86 $\pm$ 3.34 & 19.19 $\pm$ 3.30 & 19.08 $\pm$ 3.30 \\
\midrule
\multicolumn{7}{l}{\textit{NMSE (T1)}} \\
& t200 & \textbf{0.0045} $\pm$ 0.0007 & 0.0080 $\pm$ 0.0013 & 0.0046 $\pm$ 0.0010 & 0.0063 $\pm$ 0.0011 & 0.0164 $\pm$ 0.0335 \\
& t400 & \textbf{0.0029} $\pm$ 0.0005 & 0.0052 $\pm$ 0.0011 & 0.0032 $\pm$ 0.0007 & 0.0041 $\pm$ 0.0008 & 0.0126 $\pm$ 0.0186 \\
& t600 & \textbf{0.0026} $\pm$ 0.0005 & 0.0049 $\pm$ 0.0011 & 0.0031 $\pm$ 0.0007 & 0.0040 $\pm$ 0.0009 & 0.0101 $\pm$ 0.0101 \\
& t800 & \textbf{0.0026} $\pm$ 0.0005 & 0.0048 $\pm$ 0.0010 & 0.0030 $\pm$ 0.0007 & 0.0038 $\pm$ 0.0009 & 0.0058 $\pm$ 0.0058 \\
& t1000 & \textbf{0.0025} $\pm$ 0.0004 & 0.0047 $\pm$ 0.0010 & 0.0029 $\pm$ 0.0007 & 0.0036 $\pm$ 0.0008 & 0.0052 $\pm$ 0.0052 \\
\cmidrule(lr){2-7}
\multicolumn{7}{l}{\textit{NMSE (T2)}} \\
& t200 & \textbf{0.0424} $\pm$ 0.0081 & 0.0768 $\pm$ 0.0106 & 0.0540 $\pm$ 0.0122 & 0.0602 $\pm$ 0.0120 & 0.0881 $\pm$ 0.0277 \\
& t400 & \textbf{0.0274} $\pm$ 0.0054 & 0.0486 $\pm$ 0.0078 & 0.0357 $\pm$ 0.0081 & 0.0389 $\pm$ 0.0083 & 0.0513 $\pm$ 0.0189 \\
& t600 & \textbf{0.0192} $\pm$ 0.0039 & 0.0396 $\pm$ 0.0056 & 0.0274 $\pm$ 0.0062 & 0.0311 $\pm$ 0.0072 & 0.0410 $\pm$ 0.0139 \\
& t800 & \textbf{0.0160} $\pm$ 0.0032 & 0.0311 $\pm$ 0.0058 & 0.0223 $\pm$ 0.0054 & 0.0259 $\pm$ 0.0062 & 0.0306 $\pm$ 0.0095 \\
& t1000 & \textbf{0.0141} $\pm$ 0.0029 & 0.0281 $\pm$ 0.0051 & 0.0206 $\pm$ 0.0051 & 0.0238 $\pm$ 0.0058 & 0.0281 $\pm$ 0.0094 \\
\bottomrule
\end{tabular}%
}
\end{table}

%% file: tables/table_ablation.tex
\begin{table}[htbp]
\centering
\caption{Ablation study configurations of FoLaMNet architecture. IFE: Initial Fingerprint Encoder; GAST: Gate-Aware Local-Global Adaptive processor; DSFE: Deep Spatial Feature Extractor.}
\resizebox{\textwidth}{!}{%
\label{tab:ablation_configurations}
\begin{tabular}{l|ccc|l}
\toprule
\textbf{Method} & \textbf{IFE} & \textbf{GAST} & \textbf{DSFE} & \textbf{Description} \\
\midrule
\textbf{GAST-Mamba}& Mamba & \checkmark & \checkmark & Complete proposed architecture \\
A1: w/o DSFE & Mamba & \checkmark & $\times$ & Without deep spatial feature extractor \\
A2: w/o GAST & Mamba & $\times$ & \checkmark & Without GAST channel processor \\
A3: IFE only & Mamba & $\times$ & $\times$ & Only initial fingerprint encoder \\
A4: CNN encoder & CNN & \checkmark & \checkmark & CNN-based encoder instead of Mamba \\
\bottomrule
\end{tabular}
}
\end{table}

%% file: tables/table-ablation-results.tex
\begin{table}[htbp]
\centering
\caption{Ablation study results for quantitative T1 and T2 mapping reconstruction at varying timepoints. Best results are shown in bold.}
\label{tab:ablation-study}
\resizebox{\textwidth}{!}{%
\begin{tabular}{ll|ccccc}
\toprule
\multirow{2}{*}{\textbf{Parameter Map}} & \multirow{2}{*}{\textbf{Timepoint}} & \multicolumn{5}{c}{\textbf{Method}} \\
\cmidrule(lr){3-7}
& & \textbf{GAST-Mamba} & \textbf{A1} & \textbf{A2} & \textbf{A3} & \textbf{A4} \\
\midrule
\multicolumn{7}{l}{\textit{PSNR (dB)}} \\
\midrule
\multirow{5}{*}{\textbf{T1}} 
& t200 & \textbf{33.12} $\pm$ 1.09 & 32.58 $\pm$ 1.10 & 33.00 $\pm$ 1.10 & 32.63 $\pm$ 1.10 & 32.70 $\pm$ 1.09 \\
& t400 & \textbf{35.13} $\pm$ 1.17 & 34.64 $\pm$ 1.14 & 35.09 $\pm$ 1.16 & 34.64 $\pm$ 1.15 & 34.62 $\pm$ 1.13 \\
& t600 & \textbf{35.48} $\pm$ 1.15 & 34.89 $\pm$ 1.13 & 35.43 $\pm$ 1.15 & 34.92 $\pm$ 1.14 & 34.81 $\pm$ 1.10 \\
& t800 & \textbf{35.57} $\pm$ 1.14 & 35.02 $\pm$ 1.14 & \textbf{35.57} $\pm$ 1.13 & \textbf{35.57} $\pm$ 1.13 & 34.93 $\pm$ 1.09 \\
& t1000 & \textbf{35.65} $\pm$ 1.13 & 35.08 $\pm$ 1.10 & 35.57 $\pm$ 1.11 & 35.09 $\pm$ 1.10 & 35.01 $\pm$ 1.09 \\
\cmidrule(lr){2-7}
\multirow{5}{*}{\textbf{T2}} 
& t200 & \textbf{30.62} $\pm$ 1.54 & 30.46 $\pm$ 1.44 & 30.39 $\pm$ 1.54 & 30.44 $\pm$ 1.45 & 30.46 $\pm$ 1.49 \\
& t400 & \textbf{32.54} $\pm$ 1.70 & 32.47 $\pm$ 1.65 & 32.53 $\pm$ 1.68 & 32.46 $\pm$ 1.65 & 32.36 $\pm$ 1.61 \\
& t600 & 34.09 $\pm$ 1.73 & 33.95 $\pm$ 1.65 & \textbf{34.10} $\pm$ 1.72 & 33.95 $\pm$ 1.67 & 33.78 $\pm$ 1.62 \\
& t800 & 34.87 $\pm$ 1.75 & 34.68 $\pm$ 1.73 & \textbf{34.91} $\pm$ 1.74 & \textbf{34.91} $\pm$ 1.74 & 34.46 $\pm$ 1.61 \\
& t1000 & \textbf{35.43} $\pm$ 1.76 & 35.18 $\pm$ 1.75 & 35.38 $\pm$ 1.83 & 35.22 $\pm$ 1.74 & 35.08 $\pm$ 1.72 \\
\midrule
\multicolumn{7}{l}{\textit{SSIM}} \\
\midrule
\multirow{5}{*}{\textbf{T1}} 
& t200 & 0.97 $\pm$ 0.0070 & 0.96 $\pm$ 0.0080 & 0.97 $\pm$ 0.0071 & 0.96 $\pm$ 0.0079 & 0.96 $\pm$ 0.0078 \\
& t400 & 0.98 $\pm$ 0.0055 & 0.97 $\pm$ 0.0060 & 0.98 $\pm$ 0.0056 & 0.97 $\pm$ 0.0060 & 0.97 $\pm$ 0.0061 \\
& t600 & 0.98 $\pm$ 0.0053 & 0.97 $\pm$ 0.0058 & 0.98 $\pm$ 0.0054 & 0.97 $\pm$ 0.0058 & 0.97 $\pm$ 0.0059 \\
& t800 & 0.98 $\pm$ 0.0051 & 0.97 $\pm$ 0.0056 & 0.98 $\pm$ 0.0050 & 0.98 $\pm$ 0.0050 & 0.97 $\pm$ 0.0058 \\
& t1000 & 0.98 $\pm$ 0.0049 & 0.97 $\pm$ 0.0054 & 0.98 $\pm$ 0.0049 & 0.97 $\pm$ 0.0054 & 0.97 $\pm$ 0.0055 \\
\cmidrule(lr){2-7}
\multirow{5}{*}{\textbf{T2}} 
& t200 & 0.91 $\pm$ 0.0138 & 0.90 $\pm$ 0.0130 & 0.91 $\pm$ 0.0137 & 0.90 $\pm$ 0.0138 & 0.91 $\pm$ 0.0143 \\
& t400 & 0.95 $\pm$ 0.0086 & 0.95 $\pm$ 0.0087 & 0.95 $\pm$ 0.0087 & 0.95 $\pm$ 0.0088 & 0.95 $\pm$ 0.0094 \\
& t600 & 0.97 $\pm$ 0.0064 & 0.97 $\pm$ 0.0065 & 0.97 $\pm$ 0.0064 & 0.97 $\pm$ 0.0065 & 0.96 $\pm$ 0.0071 \\
& t800 & 0.97 $\pm$ 0.0052 & 0.97 $\pm$ 0.0054 & 0.97 $\pm$ 0.0051 & 0.97 $\pm$ 0.0051 & 0.97 $\pm$ 0.0059 \\
& t1000 & 0.98 $\pm$ 0.0046 & 0.98 $\pm$ 0.0047 & 0.98 $\pm$ 0.0047 & 0.98 $\pm$ 0.0048 & 0.97 $\pm$ 0.0051 \\
\midrule
\multicolumn{7}{l}{\textit{RMSE}} \\
\midrule
\multirow{5}{*}{\textbf{T1}} 
& t200 & \textbf{85.67} $\pm$ 10.03 & 91.13 $\pm$ 10.74 & 86.88 $\pm$ 10.29 & 90.66 $\pm$ 10.75 & 89.91 $\pm$ 10.54 \\
& t400 & \textbf{68.06} $\pm$ 8.53 & 71.93 $\pm$ 8.76 & 68.33 $\pm$ 8.55 & 71.95 $\pm$ 8.87 & 72.12 $\pm$ 8.82 \\
& t600 & \textbf{65.36} $\pm$ 8.10 & 69.88 $\pm$ 8.45 & 65.70 $\pm$ 8.11 & 69.68 $\pm$ 8.50 & 70.53 $\pm$ 8.42 \\
& t800 & \textbf{64.62} $\pm$ 7.90 & 68.85 $\pm$ 8.38 & 64.67 $\pm$ 7.85 & 64.67 $\pm$ 7.85 & 69.51 $\pm$ 8.21 \\
& t1000 & \textbf{64.06} $\pm$ 7.72 & 68.33 $\pm$ 8.03 & 64.63 $\pm$ 7.67 & 68.28 $\pm$ 8.02 & 68.95 $\pm$ 8.10 \\
\cmidrule(lr){2-7}
\multirow{5}{*}{\textbf{T2}} 
& t200 & \textbf{25.14} $\pm$ 4.31 & 25.55 $\pm$ 4.13 & 25.81 $\pm$ 4.43 & 25.62 $\pm$ 4.15 & 25.59 $\pm$ 4.25 \\
& t400 & \textbf{20.22} $\pm$ 3.73 & 20.36 $\pm$ 3.65 & 20.23 $\pm$ 3.67 & 20.37 $\pm$ 3.66 & 20.60 $\pm$ 3.63 \\
& t600 & 16.94 $\pm$ 3.16 & 17.18 $\pm$ 3.08 & \textbf{16.90} $\pm$ 3.15 & 17.18 $\pm$ 3.11 & 17.51 $\pm$ 3.09 \\
& t800 & 15.48 $\pm$ 2.91 & 15.81 $\pm$ 2.94 & \textbf{15.41} $\pm$ 2.87 & \textbf{15.41} $\pm$ 2.87 & 16.19 $\pm$ 2.86 \\
& t1000 & \textbf{14.52} $\pm$ 2.79 & 14.95 $\pm$ 2.85 & 14.63 $\pm$ 2.89 & 14.87 $\pm$ 2.82 & 15.10 $\pm$ 2.84 \\
\bottomrule
\end{tabular}%
}
\end{table}